\documentclass[a4paper]{amsart}
\usepackage{graphics}

\theoremstyle{definition}
\newtheorem{Def}{Definition}
\newtheorem{Rem}[Def]{Remark}
\newtheorem{Exa}[Def]{Example}
\newtheorem{Alg}[Def]{Algorithm}
\theoremstyle{plain}
\newtheorem{Lem}[Def]{Lemma}
\newtheorem{The}[Def]{Theorem}

\newtheorem{Pro}[Def]{Proposition}
\def\val#1{{\rm val}_{#1}}
\def\rep#1{{\rm rep}_{#1}}
\DeclareMathOperator{\per}{per}
\DeclareMathOperator{\aper}{aper}
\DeclareMathOperator{\uper}{uper}

\begin{document}
\title[Real numbers having ultimately periodic
representations\ldots]{Real numbers having ultimately periodic
  representations in abstract numeration systems}

\author[P. Lecomte]{P. Lecomte}
\email{plecomte@ulg.ac.be}
\author[M. Rigo]{M. Rigo\dag}
\email{M.Rigo@ulg.ac.be}
\thanks{\dag This is author is an FNRS Postdoctoral Researcher 
and also the corresponding author.}
\address[P. L. and M. R.]{\newline
Institut de Math\'ematique,\newline
Universit\'e de Li\`ege,\newline
Grande Traverse 12 (B 37),\newline
B-4000 Li\`ege,\newline
Belgium.}

\date{\today}
\begin{abstract}
  Using a genealogically ordered infinite regular language, we know
  how to represent an interval of $\mathbb{R}$. Numbers having an
  ultimately periodic representation play a special role in classical
  numeration systems. The aim of this paper is to characterize the
  numbers having an ultimately periodic representation in generalized
  systems built on a regular language. The syntactical properties of
  these words are also investigated.  Finally, we show the equivalence
  of the classical $\theta$-expansions with our generalized
  representations in some special case related to a Pisot number
  $\theta$.
\end{abstract}
\maketitle

\section{Introduction}
Enumerating the words of an infinite regular language $L$ over a
totally ordered alphabet $(\Sigma,<)$ by genealogical ordering gives a
one-to-one correspondence between $\mathbb{N}$ and $L$. This
observation was the starting point of the study of the so-called {\it
  abstract numeration systems} which are a natural generalization of
classical positional numeration systems like the Fibonacci system or
the $k$-ary system. More generally, abstract systems generalize
positional numeration systems where representations of integers are
computed by the greedy algorithm and where the set of all the
representations is a regular language \cite{Fr,Ho,Sh}. We were first
interested in the representation of non-negative integers and in the
syntactical properties of sets of representations \cite{LR2,Ri}.

In \cite{LR} we extended these systems to the representation of real
numbers. Mimicking the case of positional systems, a real number $x$
is represented by an infinite word $w$ which is the limit of a
sequence $(w_n)_{n\in\mathbb{N}}$ of words in $L$. Each finite word
$w_n$ of the sequence corresponding to a numerical approximation of
$x$, the longer the common prefix of $w$ and $w_n$ is, the more
accurate the approximation of $x$ is. The ability of representing not
only integers but also real numbers in abstract systems lead to
several applications and generalizations like the study of the
asymptotic properties of summatory functions of additive functions
like the ``sum-of-digits'' function \cite{GR} or the description of 
properties of generalized adding machines, i.e., ``odometers''
\cite{GLR}.

Real numbers having an ultimately periodic representations are of
particular interest. First, from the point of view of computational
aspects, the amount of data needed to store completely such a number
is finite (the information is given exactly by the aperiodic part and
one period). Next, we shall see that the set of ultimately periodic
representations is dense in the set of all the representations, so
studying this subset of representations is relevant when dealing with
approximations of real numbers. As an example, for the $k$-ary system,
it is well-known that the numbers having an ultimately periodic
representation is $\mathbb{Q}$ which is dense in $\mathbb{R}$ and
therefore various number-theoretic problems concerning rational
approximations of real numbers can arise. Finally, for classical
systems (more precisely, for $\beta$-expansions when $\beta$ is a
Pisot number) the set of real numbers having an ultimately periodic
representation is exactly the field-extension $\mathbb{Q}(\beta)$
\cite{KS}. To be able to represent real numbers in a generalized
numeration system, we consider some assumptions about the counting
function of the language, namely $\mathbf{u}(n):=\#(L\cap\Sigma^n)\sim
P (n) \theta^n$ for some polynomial $P$ and $\theta>1$. Therefore the
problem of relating this number $\theta$ to the set of numbers having
an ultimately periodic representation clearly appears in the case of
abstract systems.

This paper has the following organization.  First we recall
definitions and notation about abstract numeration systems and the
representation of real numbers. Next we recall the general assumptions
we consider when dealing with the representation of real numbers. As
stated before, these assumptions are related to the asymptotic
behavior of the counting functions of the languages accepted from the
different states of the minimal automaton of $L$. The reader could 
already note that we have slightly simplified the presentation given
in \cite{LR}. The aim of Sections \ref{sec:1} and \ref{sec:2} is to
give a summary of the relevant facts given in \cite{LR2,LR,Ri}. 

In Section \ref{sec:sec3} we study the syntactical properties of the
ultimately periodic representations. We show that the corresponding
language of infinite words is $\omega$-rational. This section has an
automata theory flavor and can be read separately from the rest of the
paper.

In Section \ref{sec:sec4}, we obtain formulas for computing
effectively the numerical value of an ultimately periodic
representation. Moreover we show that the language made up of the
ultimately periodic representations is dense in the set of all the
representations. In \cite{LR}, it is explained that for an abstract
system built upon an arbitrary regular language $L$, a real number can
have one, a finite number or even an infinite number of
representations and the situation can be completely determined from
the language $L$ (actually, from the asymptotic behavior of the
counting functions associated to the different states).  In Section
\ref{sec:sec5}, we show how to modify the language $L$ to obtain a new
numeration system having exactly the same representations except that
in this new system a number has at most two representations. Roughly
speaking, we remove from the minimal automaton of $L$ the useless
states which are giving redundant representations. 

In Section \ref{sec:sec6}, we use some intervals $I_w$ (a real number
$x$ belongs to $I_w$ if $x$ has a representation having $w$ as prefix)
to obtain a characterization of the real numbers having an ultimately
periodic representation. From the ideas given in this result and its
proof, we derive two algorithms for computing the representation of an
arbitrary real number. These algorithms can be viewed as a
generalization of the greedy algorithm used to compute
$\beta$-expansions \cite{Pa} and rely on the use of some affine
functions completely defined by the minimal automaton of the language.
We also present a dynamical system built upon those affine functions,
the points having an ultimately periodic orbit in this dynamical
system being exactly the real numbers having an ultimately periodic
representation (this system is a generalization of the intervals
exchange transformation \cite{iex1,iex2}). In Section \ref{sec:sec7},
thanks to our algorithm of representation, we obtain another
characterization of the real numbers having an ultimately periodic
representation, these numbers are the fixed points of composition of
some affine functions. Moreover, this composition can actually be
viewed as a word belonging to a regular language over a finite
alphabet of functions.

In the last section, we consider a Pisot number $\theta$.  To this
number, corresponds a unique linear Bertrand numeration system
\cite{BH}. If $L$ is the language of representations of the integers
in this latter Bertrand system then the representations of the real
numbers in the abstract system built upon $L$ and the classical
$\theta$-developments are the same. So thanks to a famous result of
Klaus Schmidt, in this particular case, we know precisely the
structure of the set of real numbers having an ultimately periodic
representation. This set is $\mathbb{Q}(\theta)$.

\section{Preliminaries}\label{sec:1}
Let us precise notation and definitions. Let $\Sigma$ be a finite
alphabet. We denote by $\Sigma^*$ the free monoid generated by
$\Sigma$ with identity $\varepsilon$. Let $L$ be an infinite regular
language and ${\mathcal M}_L=(Q,q_0,\Sigma,\delta,F)$ be its minimal
automaton having $Q$ as set of states, $q_0$ as initial state, $F$ as
set of final states.  The transition function $\delta:Q\times
\Sigma\to Q$ of this automaton is naturally extended to
$Q\times\Sigma^*$ and we often write $q.w$ as a shorthand for
$\delta(q,w)$, $q\in Q$, $w\in\Sigma^*$. (For more about automata
theory see for instance \cite{E}.) If $q\in Q$, we denote by $L_q$ the
language accepted in ${\mathcal M}_L$ from the state $q$, i.e.,
$$L_q=\{w\in\Sigma^*\mid q.w \in F\}.$$
In particular, $L_{q_0}=L$. In
this paper, we shall extensively use the following linear recurrent
sequences defined for each $q\in Q$ by 
$$\mathbf{u}_q(n)=\#(L_q\cap\Sigma^n),\quad
\mathbf{v}_q(n)=\#(L_q\cap\Sigma^{\le n}).$$
Since the initial state
$q_0$ plays a special role, if $q=q_0$ then we simply write
$\mathbf{u}(n)$ and $\mathbf{v}(n)$ (in the literature,
$\mathbf{u}(n)$ is often called the growth function, the counting
function or even the complexity of the language $L$).

Let $(\Sigma,<)$ be a totally ordered alphabet. The {\it genealogical
  ordering} (or radix ordering) is defined as follows. Let $u$, $v$ be
two words over $\Sigma$, $u<_{gen}v$ if $|u|<|v|$ or if the words have
the same length and $u$ is lexicographically less than $v$ (the
lexicographic ordering is the usual order of the dictionary). If the
context is clear, we write $u<v$ instead of $u<_{gen}v$.

In \cite{LR2} we introduced numeration systems generalizing classical
numeration systems in which the set of representations of all the
integers is a regular language. An {\it abstract numeration system} is
a triple $S=(L,\Sigma,<)$ where $L$ is infinite regular language over
the totally ordered alphabet $(\Sigma,<)$. The genealogical ordering
of $L$ induced by the ordering of $\Sigma$ gives a one-to-one
correspondence between $\mathbb{N}$ and $L$.  If $w\in L$, we denote
by $\val{S}(w)$ the position of $w$ in the genealogically ordered
language $L$ (positions are counted from zero).  The number
$\val{S}(w)$ is said to be the {\it numerical value} of $w$.
Conversely, let $n\in\mathbb{N}$, if $w$ is the $(n+1)$th word in the
genealogically ordered language $L$, then $w$ is the {\it
  $S$-representation} of $n$ and is denoted by $\rep{S}(n)$ (so
$\rep{S}=\val{S}^{-1}$). In particular, these abstract systems
generalize the well known class of positional linear numeration
systems built upon a Pisot number \cite{BH,FS}. These latter systems
are constructed on a strictly increasing sequence of integers
satisfying a linear recurrence relation whose characteristic
polynomial is the minimal polynomial of a Pisot number (a Pisot number
is an algebraic integer $\theta>1$ whose Galois conjugates have
modulus less than one).

A numeration system has to be able to represent not only integers but
also real numbers. So in \cite{LR} we described how to obtain the
representations of the elements belonging to an interval of real
numbers of the form $[1/\theta,1]$ in an abstract numeration system
(and therefore using some conventions we can represent $[0,1]$). Let
$k\in\mathbb{N}\setminus\{0,1\}$. In the $k$-ary numeration system, a
real number $x\in (0,1)$ is represented by an infinite word
$w=w_0w_1w_2\cdots$. On the one hand, we have finite prefixes
$w_0\cdots w_{n-1}$ of $w$ converging to the infinite word $w$. On the
other hand, each prefix $w_0\cdots w_{n-1}$ gives rise to a numerical
approximation
\begin{equation}\label{eq:11}
\frac{\sum_{i=0}^{n-1} w_{i} k^{n-i-1}}{k^n}
\end{equation}
and the sequence of these numerical approximations is converging to
the real number $x$.  Actually, the numerator in \eqref{eq:11} is the
numerical value in base $k$ of $w_0\cdots w_{n-1}$ and the denominator
is the number of words of length at most $n$ in the language
$\{\varepsilon\}\cup\{1,\ldots ,k-1\}\{0,\ldots ,k-1\}^*$ associated
to the $k$-ary system. Having in mind these two kinds of convergence,
we proceed in the same way for an abstract system built upon a regular
language $L$ and consider sequences of words in $L$ converging to an
infinite word. Mimicking the formula \eqref{eq:11}, the numerical
approximation given by a word $w\in L\cap \Sigma^n$ is
$$\frac{\val{S}(w)}{\mathbf{v}(n)}.$$
Under suitable assumptions, a sequence of
numerical approximations is convergent whenever the corresponding
sequence of words is convergent \cite{LR}.

\section{Framework for the representation of real numbers}\label{sec:2}
To represent real numbers in an abstract numeration system, we consider
converging sequences of words in $L$. So we introduce the following
notation
$${\mathcal L}_\infty=\{w\in\Sigma^\omega \mid \exists
(w_n)_{n\in\mathbb{N}}\in L^{\mathbb{N}} : \lim_{n\to\infty}w_n=
w\}.$$
This set dedicated to be the set of the representations of the
considered real numbers (the interval of real numbers that we are able
to represent will be make explicit soon). Therefore ${\mathcal
  L}_\infty$ has to be uncountable (because we want to represent an
interval of real numbers which is uncountable).  In \cite{LR}, to be
able to prove the convergence of the numerical approximations, we
considered the following work hypothesis concerning the asymptotic
behavior of the sequences $\mathbf{u}_q(n)$'s.

\medskip
\noindent
{\bf Hypothesis.} The set ${\mathcal L}_{\infty}$ is uncountable and 
for all states $q$ of ${\mathcal M}_L$, either
\begin{itemize}
\item[(i)] $\exists N_q \in \mathbb{N}:$ $\forall n>N_q$, ${\mathbf
    u}_q(n)=0$ or
\item[(ii)] there exist $\theta_q \ge 1$, $P_q(x) \in \mathbb{R} [x]$
  and $b_q>0$ such that
$$\lim_{n\to \infty} \frac{{\mathbf u}_{q}(n)}{P_q(n) \theta_q^n} = b_q.$$
\end{itemize}
Since ${\mathcal L}_{\infty}$ is uncountable, it can be shown that
the language $L$ has an exponential growth and therefore 
$\theta=\theta_{q_0}>1$. 

By choosing the coefficient of the dominant
term in $P_{q_0}$ (or in the same way by replacing $P_{q_0}$ with
$P_{q_0}/b_{q_0}$), we may assume in what follows that
$$\lim_{n\to \infty} \frac{{\mathbf u}_{q_0}(n)}{P_{q_0}(n) \theta^n} = 1.$$
For all states $q$ of ${\mathcal M}_L$, the following limit exists
$$\lim_{n\to \infty} \frac{{\mathbf u}_{q}(n)}{P_{q_0}(n) \theta^n}$$
and we denote by $a_q\ge 0$ its value ($a_{q_0}=1$). For details, see
\cite{LR}.
\begin{Rem}
If a state $q$ is such that $d(P_q)<d(P_{q_0})$ or $\theta_q<\theta$ 
then $a_q=0$.
\end{Rem}

\begin{Pro}\label{pro:formule}{\rm \cite[Corollary 7]{LR}} 
  If $(w_n)_{n\in\mathbb{N}} \in L^{\mathbb{N}}$ is converging to an
  infinite word $w\in\mathcal{L}_\infty$ then
  $$\lim_{n\to \infty}\frac{\val{S}(w_n)}{{\mathbf v}(\vert w_n
      \vert)} = \frac{\theta -1}{\theta^2} \sum_{q\in Q} a_q\,
  \sum_{j=0}^\infty \beta_{q,j} \, \theta^{-j}=x$$
  where the
  coefficients $\beta_{q,j}$ depends only on $w$.
\end{Pro}
In this latter proposition, the infinite word $w$ is said to be a {\it
  representation} of the real number $x$. In the same way, $x$ is said
to be the {\it numerical value} of $w$. Each real number in
$[1/\theta,1]$ has at least one representation in
$\mathcal{L}_\infty$. Conversely, each infinite word in
$\mathcal{L}_\infty$ is the representation of a unique number in
$[1/\theta,1]$.

\begin{Rem}\label{rem:trans}
  We know precisely what are the coefficients $\beta_{q,j}$'s
  introduced in Proposition \ref{pro:formule}. If the infinite word
  $w$ is written $w_0w_1\cdots $ then for all states $q$
$$\beta_{q,j}=\#\{\sigma<w_j\mid q_0.w_0\cdots w_{j-1}\sigma =
q\}+\delta_{q,q_0}$$
where $\delta$ is the Kronecker's symbol. As
noticed in \cite{GR}, those coefficients can be computed by a
transducer $\mathcal{T}$ built upon $\mathcal{M}_L$. If
$Q=\{q_0,q_1,\ldots ,q_r\}$ and if in $\mathcal{M}_L$,
$p.\sigma=p'$ then in $\mathcal{T}$, the directed edge between
$p$ and $p'$ is labeled by $(\sigma,n_0,n_1,\ldots ,n_r)$ where
$$n_i=\#\{\tau<\sigma\mid p.\tau=q_i\}+\delta_{i,0}.$$
The
reading in $\mathcal{T}$ of the $n$th letter of a word
$w=w_0w_1\cdots$ gives the $(r+1)$-uple $(\beta_{q_0,n},\ldots
,\beta_{q_r,n})$ corresponding to $w$, $n\ge 0$.
\end{Rem}

\begin{Exa}
  Consider the language made up of the words containing an even number
  of $a$'s (we assume that $a<b$). The transducer computing
  simultaneously the coefficients $\beta_{q_0,n}$ and $\beta_{q_1,n}$
  is given in Figure \ref{fig:autom3}.
  \begin{figure}[h!tbp]
    \begin{center}
      \includegraphics{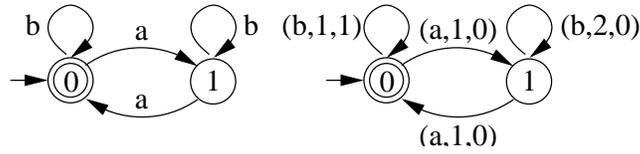}
      \caption{An automaton and the corresponding 
transducer computing $(\beta_{q_0,j},\beta_{q_1,j})_{j\in\mathbb{N}}$.}
      \label{fig:autom3}
    \end{center}
  \end{figure}
\end{Exa}

\section{Syntactical properties of the periods in $\mathcal{L}_\infty$}\label{sec:sec3}

In this section, we are interested in the syntactical properties of the
ultimately periodic representations of real numbers. Namely, if
$uv^\omega$ is an element in $\mathcal{L}_\infty$ then what can be
said about the syntax of $u$ or $v$ ? Are the words $u$ and $v$
related in some way to $L$ ?

\begin{Def} 
  Let $X$ be a set of infinite words. The set of the periods of the
  ultimately periodic words in $X$ is denoted 
  ${\per}(X)$ and is defined by
  $$v \in {\per}(X) \Leftrightarrow \exists u\in\Sigma^*: uv^\omega
  \in X .$$
  In the same way, we can also define the set ${\aper}(X)$
  of the aperiodic parts of the ultimately periodic words in $X$,
  $$u \in {\aper}(X) \Leftrightarrow \exists v\in\Sigma^+: uv^\omega
  \in X.$$
  
\end{Def}

\begin{Exa} 
  The periodic and aperiodic parts of a word are not necessarily
  unique. Consider the word
  $$w=aba(ab)^\omega,$$
  we have ${\per}(\{w\})=\{(ab)^i\mid i>0\}$ and
  ${\aper}(\{w\})$ contains any prefix of $w$ of length at least $3$.
  Nevertheless, the minimal aperiodic prefix is $aba$ and to this
  prefix is corresponding the period $ab$ of minimal length.
\end{Exa}

The first aim of this section is to show that ${\per}({\mathcal
  L}_\infty)\subseteq \Sigma^*$ is a regular language. Next we shall
see that ${\aper}({\mathcal L}_\infty)\subseteq \Sigma^*$ is also
regular.

In the following lemma, we are interested in the states reached in
$\mathcal{M}_L$ when reading an ultimately periodic word.
\begin{Lem}\label{lem} 
Let $(x_n)_{n\in\mathbb{N}}=uv^\omega$ be an ultimately periodic word of ${\mathcal L}_\infty$, the word
  $$\xi=(q_0,x_0) (q_0.x_0,x_1) (q_0.x_0x_1,x_2) \cdots (q_0.x_0\cdots
  x_n,x_{n+1})\cdots \in (Q\times \Sigma)^\omega$$
  is ultimately
  periodic of period $t|v|$ for some $t\le \#Q$.
\end{Lem}

\begin{proof} 
  We use the same kind of reasoning as in the proof of the classical
  pumping lemma (\cite[Lemma 4.1]{Yu}). If $\# Q=r$ then, for any
  $q\in Q$, at least two states of the following list of $r+1$ states
  are the same
  $$q, q.v, \ldots ,q.v^r.$$
  So for $q=q_0.u$, there exist $i$, $j$,
  $0\le i<j \le r$, such that $q'=q.v^i=q.v^j$. Thus in
  $\mathcal{M}_L$ after reading $v^{j-i}$ from this state $q'$, we are
  back in $q'$ and we have still to read $v^\omega$. The deterministic
  behavior of $\mathcal{M}_L$ leads to the conclusion. Notice that the
  period of $\xi$ is bounded by $(j-i)\, |v|\le \# Q\, |v|$.
\end{proof}

We now present the construction of an automaton $\mathfrak{M}$ that
will be used to show that ${\per}({\mathcal L}_\infty)$ is regular.

\begin{Def}\label{def:cycle}
  Let us define a set ${\mathcal C}\subseteq
  2^Q\setminus\{\emptyset\}$.  A set $C=\{p_1,\ldots ,p_k\}$ of states
  belongs to ${\mathcal C}$ if and only if the following two
  conditions are satisfied
\begin{enumerate}
\item $C$ is a {\it cycle} in ${\mathcal M}_L$ : there exists a word 
  $w=w_1\cdots w_\ell\in \Sigma^\ell$, $\ell \ge k$, such that 
  \begin{itemize}
  \item $p_1.w=p_1$,
  \item $\forall i\le\ell$, $p_1.w_1\cdots w_i\in C$,
  \item $\forall i\le k$, $\exists j\le \ell$: $p_1.w_1\cdots
    w_j=p_i$.
  \end{itemize}
\item $C$ is {\it coaccessible} : there exist $p_i\in C$ and
  $w\in\Sigma^*$ such that $p_i.w\in F$.
\end{enumerate}
\end{Def}

\begin{Rem} 
  Let $C\in\mathcal{C}$. The set $C$ is also accessible. Indeed, since
  $\mathcal{M}_L$ is minimal, it is accessible.  So for each state
  $p\in C$, there exists a word $w$ such that $q_0.w=p$. 
  
  Another observation is the following. In Definition \ref{def:cycle},
  the cycle given by the word $w=w_1\cdots w_\ell$ is not necessarily
  an Hamiltonian circuit.
\end{Rem}

\medskip
\noindent
It is clear that ${\mathcal C}$ is ordered by inclusion. We denote by
$C_1,\ldots ,C_t$ the maximal elements of $({\mathcal C},\subseteq)$.
(As a consequence of the maximality, for $i\neq j$, $C_i\cap
C_j=\emptyset$. Indeed, if a state belongs to $C_i\cap C_j$ then
$C_i\cup C_j$ belongs to $\mathcal{C}$ because we can find a longer
cycle in $\mathcal{M}_L$ and therefore neither $C_i$ nor $C_j$ is
maximal.) 

Let $i\in \{1,\ldots ,t\}$, $C_i=\{p^{(i)}_1,\ldots ,p^{(i)}_{r_i}\}$ ($r_i\ge
1$) and $j\in\{1,\ldots, r_i\}$.  For each such indices $i$ and $j$, we
define a NFA $\mathcal{M}_{i,j}$ in the following way,
$$\mathcal{M}_{i,j}=(C_i,p^{(i)}_j,\Sigma,\delta\vert_{C_i\times
  \Sigma\times C_i},\{p^{(i)}_j\}),$$
the set of states is $C_i$, the
initial state is $p^{(i)}_j$ and the transition relation is the restriction
of the transition function $\delta$ of $\mathcal{M}_L$ to the states
belonging to $C_i$ : $\delta(p,\sigma)=q$ with $p,q\in C_i$ iff
$(p,\sigma,q)$ belongs to $\delta\vert_{C_i\times \Sigma\times C_i}$.
The state $p^{(i)}_j$ is the unique final state of $\mathcal{M}_{i,j}$.
(Observe that the automaton $\mathcal{M}_{i,j}$ is non-deterministic
only because the function $\delta$ is not necessarily complete.)

Let $\mathfrak{M}$ be the NFA obtained as the union of the different
$\mathcal{M}_{i,j}$'s. In this construction we assume that the sets of
states of two distinct automata $\mathcal{M}_{i,j}$'s are disjoint. To
obtain the union, we only have to consider a set of initial states
instead of a single one.

\begin{Exa} 
  Let us consider the trim minimal automaton depicted in Figure
  \ref{fig:autom} (the sink has not been represented, this state is
  never coaccessible so it never belong to a set in $\mathcal{C}$).
    \begin{figure}[h!tbp]
      \begin{center}
        \includegraphics{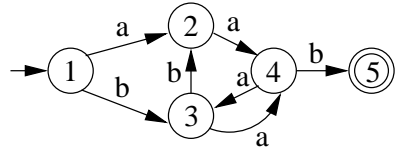}
        \caption{A minimal automaton $\mathcal{M}_L$ (without sink).}
        \label{fig:autom}
      \end{center}
    \end{figure}
    Here $\mathcal{C}=\{\{3,4\},\{2,3,4\}\}$ and we have a single
    maximal element $C_1=\{2,3,4\}$ in $(\mathcal{C},\subseteq)$.  The
    corresponding NFA $\mathfrak{M}$ is given in Figure
    \ref{fig:autom2}. (We have three initial states.) With our
    notation, this automaton is built (from left to right in Figure
    \ref{fig:autom2}) upon the automata $\mathcal{M}_{1,2}$,
    $\mathcal{M}_{1,4}$ and $\mathcal{M}_{1,3}$.
\begin{figure}[h!tbp]
      \begin{center}
        \includegraphics{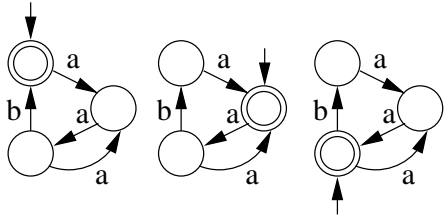}
        \caption{The corresponding NFA $\mathfrak{M}$ with 
three initial states.}
        \label{fig:autom2}
      \end{center}
    \end{figure}
\end{Exa}

Let $k>0$. Recall that a word $u$ belongs to the $k$th-root
$\sqrt[k]{L}$ of a language $L$ if and only if $u^k$ belongs to $L$.
If $\mathcal{M}$ is an automaton (deterministic or not), the language
accepted by $\mathcal{M}$ is denoted $L(\mathcal{M})$.

\begin{Pro}\label{pro:1} We have
$${\per}({\mathcal L}_\infty)=\bigcup_{k=1}^{\# Q}\sqrt[k]{L(\mathfrak{M})}.$$
In particular ${\per}(\mathcal{L}_\infty)$ is regular.
\end{Pro}

\begin{proof} Let $w^k\in L(\mathfrak{M})$ for some $k\le \# Q$. 
  This means that $w^k$ is accepted by some $\mathcal{M}_{i,j}$ and
  $p^{(i)}_j.w^k=p^{(i)}_j$.  By definition of $\mathcal{C}$ and
  $\mathcal{M}_{i,j}$, it is clear that in $\mathcal{M}_L$, there
  exist $x$ and $y$ in $\Sigma^*$ such that $q_0.x=p^{(i)}_j$ and
  $p^{(i)}_j.y\in F$. Therefore, we have
  $$\forall n\in\mathbb{N},\ x w^{kn} y \in L$$
  and $xw^\omega \in
  {\mathcal L}_\infty$ since $x w^{kn} y \to xw^\omega$ if $n$ tends
  to infinity. So $w$ is a period of an ultimately periodic word in
  $\mathcal{L}_\infty$.

Let $w$ be an element of ${\per}({\mathcal L}_\infty)$. There exists
an ultimately periodic word $u=v w^\omega=u_0u_1u_2\cdots \in {\mathcal
  L}_\infty$ having $w$ as period. By Lemma \ref{lem}
  $$\xi=(q_0,u_0) (q_0.u_0,u_1) (q_0.u_0u_1,u_2) \cdots (q_0.u_0\cdots u_n,u_{n+1})\cdots $$
is ultimately periodic of period $t |w|$ for some $t\le \# Q$. We can write $\xi=\alpha \beta^\omega$.
\begin{figure}[h!tbp]
  \begin{center}
    \includegraphics{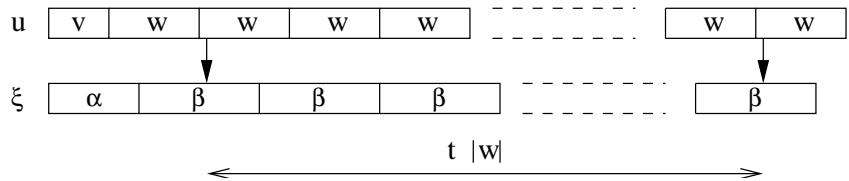}
    \caption{The periods.}
    \label{fig:period}
  \end{center}
\end{figure}
After an initial mess $\alpha$, the periodic part $\beta^\omega$ of
$\xi$ begins. In this latter part, there is a position $(q,\sigma)$
corresponding to the first reading of the beginning of $w$. (Imagine
that the words $u$ and $\xi$ are written on two tapes which are read
simultaneously by a single head one element at a time, so to the $n$th
letter of $u$ corresponds exactly the $n$th element of $\xi$. This
situation is depicted schematically in Figure \ref{fig:period}.) By
periodicity of $\xi$, after reading $t |w|$ letters, we are again in
the same position $(q,\sigma)$ and by periodicity of
$(u_n)_{n\in\mathbb{N}}$, we are ready to read the beginning of a new
occurrence of $w$. This means that we have in $\mathcal{M}_L$ a cycle
containing $q$ and such that $q.w^t=q$. So clearly $w^t$ is accepted
by some previously defined automaton $\mathcal{M}_{i,j}$ having $q$ as
initial state. In other words, since $t\le \# Q$, $$w\in
\sqrt[t]{L(\mathfrak{M})}\subseteq \bigcup_{k=1}^{\#
  Q}\sqrt[k]{L(\mathfrak{M})}.$$
We can conclude, applying Lemma \ref{lem:2} below.
\end{proof}

\begin{Lem}\label{lem:2} 
  If $L$ is a regular language then for $k\ge 1$, $\sqrt[k]{L}$ is
  also regular.
\end{Lem}
For the sake of completeness, we recall the proof of Lemma
\ref{lem:2}.
\begin{proof} 
If $q=q_0.w$ is a state of $\mathcal{M}_L$ then 
$$q=w^{-1}.L=\{v\in\Sigma^*\mid wv\in L\}$$
is a regular language. For the properties of the
minimal automaton of a language, we refer to \cite[III.5]{E}.
We define the language
$$\mathcal{S}(q)=\{m\in\Sigma^*\mid q=m^{-1}.L\}.$$
This language is
also regular. If $Q$ is the finite set of states of $\mathcal{M}_L$,
then for $k>0$ the following formula holds
$$\sqrt[k+1]{L}=\bigcup_{q\in Q} (\mathcal{S}(q) \cap \sqrt[k]{q}).$$ 
Hence we obtain the conclusion by using an easy induction argument. 
\end{proof}

We can now show that ${\aper}({\mathcal L}_\infty)$ is regular.

If $q$ is a state belonging to a maximal element $C_i$ of
$\mathcal{C}$ then we denote by $\mathcal{M}_q$ the DFA
$\mathcal{M}_q=(Q,q_0,\Sigma,\delta,\{q\})$ built upon the minimal
automaton of $L$ where the set of final states is replaced by $\{q\}$.
The following proposition is obvious.
\begin{Pro}
If the maximal elements of $\mathcal{C}$ are $C_1,\ldots,C_t$ then
$${\aper}({\mathcal L}_\infty)=\bigcup_{q\in C_1\cup\cdots\cup C_t}
L(\mathcal{M}_q)$$
and this language is therefore regular.
\end{Pro}

\begin{Pro} The language ${\uper}({\mathcal L}_\infty)\subset\Sigma^\omega$ of the ultimately periodic words in $\mathcal{L}_\infty$ is $\omega$-rational.
\end{Pro}
\begin{proof} If the maximal elements of 
  $\mathcal{C}$ are $C_1,\ldots,C_t$ then, using the previous notation
  for the automata $\mathcal{M}_q$, we have
  $${\uper}({\mathcal L}_\infty)= \bigcup_{i=1}^t\left[\bigcup_{q\in
      C_i} L(\mathcal{M}_q) \cdot \left(\bigcup_{k=1}^{\# C_i}
      \sqrt[k]{L(\mathcal{M}_{i,q})} \right)^\omega\right]$$
  where
  $\mathcal{M}_{i,q}$ is the NFA having $C_i$ as set of states and $q$
  as initial and final state. In this formula, the dot ``$\cdot$''
  represents the concatenation.  Hence the conclusion, using classical
  results on $\omega$-rational languages \cite{WT}.
\end{proof}

\section{Computing the values of ultimately periodic representations}\label{sec:sec4}
Thanks to Proposition \ref{pro:formule}, the numerical value of an
ultimately periodic word in $\mathcal{L}_\infty$ can be easily
computed.  As a consequence of Remark \ref{rem:trans}, if $w$ is an
infinite ultimately periodic word with a period of length $\ell$ then
the corresponding sequence $(\beta_{q,n})_{n\in\mathbb{N}}$ is
ultimately periodic with a period of length bounded by $\#Q.\ell$.

The following lemma is obvious.

\begin{Lem}\label{lem:per} If $(\alpha_n)_{n\in\mathbb{N}}$ is an ultimately periodic sequence of real numbers of the form $\alpha_0,\ldots,\alpha_{r-1},(\alpha_r,\ldots,\alpha_{r+p-1})^\omega$ then
  $$\sum_{j=0}^\infty \alpha_j
  \theta^{-j}=\sum_{j=0}^{r-1}\alpha_j\theta^{-j}
  +\frac{\theta^p}{\theta^p-1}\sum_{j=r}^{r+p-1}\alpha_j\theta^{-j}.$$
\end{Lem}

If $(\beta_{q,n})_{n\in\mathbb{N}}$ is ultimately periodic, we denote
by $r_q$ the minimal length of its aperiodic part and by $p_q$ the
length of the corresponding minimal period of the sequence. As a
consequence of Proposition \ref{pro:formule} and Lemma \ref{lem:per},
if $(w_n)_{n\in\mathbb{N}} \in L^{\mathbb{N}}$ is converging to an
ultimately periodic word $w$ corresponding to ultimately periodic
sequences $(\beta_{q,n})_{n\in\mathbb{N}}$, then we have
\begin{equation}\label{eq:1}
\lim_{n\to \infty}\frac{\val{S}(w_n)}{{\mathbf v}(\vert w_n \vert)} = \frac{\theta 
-1}{\theta^2} \sum_{q\in Q} a_q\, \left(
\sum_{j=0}^{r_q-1}\beta_{q,j}\theta^{-j} 
+\frac{\theta^{p_q}}{\theta^{p_q}-1}
\sum_{j=r_q}^{r_q+p_q-1}\beta_{q,j}\theta^{-j} \right).
\end{equation}
If $r=\max_{q\in Q} r_q$ and $p={\rm lcm}_{q\in Q}\ p_q$ then 
\begin{equation}\label{eq:2}
\lim_{n\to \infty}\frac{\val{S}(w_n)}{{\mathbf v}(\vert w_n \vert)} 
= \frac{\theta -1}{\theta^2}  \left(
\sum_{j=0}^{r-1} (\sum_{q\in Q} a_q\, \beta_{q,j})\, \theta^{-j} 
+\frac{\theta^{p}}{\theta^{p}-1}
\sum_{j=r}^{r+p-1}(\sum_{q\in Q} a_q\, \beta_{q,j})
\, \theta^{-j} \right).
\end{equation}

\begin{Rem}  
  The coefficients $\sum_{q\in Q} a_q\, \beta_{q,j}$ could also be
  computed by a transducer built in a similar way as $\mathcal{T}$.
\end{Rem}

With formulas \eqref{eq:1} or \eqref{eq:2}, we can compute easily the
real number represented by an ultimately periodic word. 

Moreover, $\theta$ is an eigenvalue of the adjacency matrix $A$ of
$\mathcal{M}_L$ and $\overrightarrow{a}=(a_{q_0},a_{q_1},\ldots
,a_{q_r})$ is one of its eigenvectors (recall that if $p,q\in Q$ then
the {\it adjacency matrix} is defined by
$A_{p,q}=\#\{\sigma\in\Sigma\mid p.\sigma=q\}$). For $n\ge 1$
and $p\in Q$, it is obvious that
$$\mathbf{u}_p(n)=\sum_{q\in Q}A_{p,q} \mathbf{u}_q(n-1).$$
Dividing
both sides by $P_{q_0}(n)\theta^n$ and letting $n$ tend to infinity,
we obtain
$$a_p=\frac{1}{\theta} \sum_{q\in Q}A_{p,q} a_q.$$
So
$A\overrightarrow{a}=\theta \overrightarrow{a}$. This latter
observation could be useful to determine in a practical way the value of
$\overrightarrow{a}$ (remember that we have chosen $a_{q_0}$ to be
equal to $1$).

To conclude this section, we show that the set of ultimately periodic
words in $\mathcal{L}_\infty$ is dense in $\mathcal{L}_\infty$. This
can be related to the classical fact that for the base $10$ system,
$\mathbb{Q}$ is exactly the set of numbers having an ultimately
periodic representation and $\mathbb{Q}$ is dense in $\mathbb{R}$.
\begin{Pro} 
  The set ${\uper}({\mathcal L}_\infty)\subset\Sigma^\omega$ of the
  ultimately periodic words in $\mathcal{L}_\infty$ is dense in
  $\mathcal{L}_\infty$
\end{Pro}
\begin{proof} 
  Let $w\in \mathcal{L}_\infty$. By definition of the set, there
  exists a sequence $(w_n)_{n\in\mathbb{N}}$ of words in $L$ such that
  $w_n\to w$. For any $\ell>0$ there exist $N$ such that $w$ and $w_N$
  have a common prefix of length $\ell$ and $|w_N|\ge\ell+\# Q$. When
  reading the suffix of length $\# Q$ of $w_N$ in $\mathcal{M}_L$ we
  go at least twice through the same state $q$ and let $u$ be the
  corresponding factor of $w_N$ such that $q.u=q$. Therefore, $w_N$
  can be written $xuz$ with $|x|\ge l$ and it is clear that $xu^nz\in
  L$ for all $n\ge 1$.  So $xu^\omega\in \mathcal{L}_\infty$ and has a
  prefix of length $\ell$ in common with $w$. We can therefore build a
  sequence of ultimately periodic words converging to $w$.
\end{proof}

\section{Simplifying the language}\label{sec:sec5}
In the first part of this section, we explain how a real number can
have more than one representation and even an infinite number of
representations. Next, we explain, how we can slightly change the
language to avoid this situation of having an infinite number of
representations but without altering the other representations.

In \cite{LR}, we gave a partition of the interval $[1/\theta,1]$ into
intervals $I_w$. These intervals will play a central role in what
follows so let us recall their definition. First consider the $k$-ary
system. In this system, the representation of a real number
$x\in[1/10,1]$ has a prefix $w=w_0\cdots w_n$ ($w_0\neq 0$) if $x$
belongs to the interval
$$I_w=\left[\frac{\sum_{i=0}^{n-1} w_{i}
    k^{n-i-1}}{k^n},\frac{1+\sum_{i=0}^{n-1} w_{i}
    k^{n-i-1}}{k^n}\right]$$
Observe that the endpoints of the
intervals $I_w$ are the only numbers having two representations. For
instance, if $k=10$ then $2/10$ can be written $0,1999\cdots$ and
$2/10$ is the upper bound of $I_1$ but it can also be written
$0,2000\cdots$ and is the lower bound of $I_2$. For an abstract
numeration system, we have the following definition.

\begin{Def}
  A real number $x\in [1/\theta,1]$ belongs to $I_w$ if there exist a
  representation of $x$ having $w$ as prefix.
\end{Def}
 
For an arbitrary regular language $L$, the set $\mathcal{L}_\infty$
can contains an infinite number of words having $w$ as prefix even if
the length of the interval $I_w$ is zero. In this situation, all the
elements of $\mathcal{L}_\infty$ having $w$ as prefix are representing
the same real number $x$ and $I_w=[x,x]$. Therefore $x$ has an
infinite number of representations.  To avoid this situation, we
proceed as follows.

We can only consider the states $q$ such that $a_q>0$.
We have the following rules
\begin{itemize}
\item If $a_q\neq 0$ and there exists $w$ such that $p.w=q$, then $a_p\neq 0$.
\item If $a_q=0$ and there exists $w$ such that $q.w=p$, then $a_p=0$.
\end{itemize}
Indeed, in the first case, if $p.w=q$ then $\mathbf{u}_n(p)\ge
\mathbf{u}_{n-|w|}(q)$. In the second case, if $q.w=p$ then $\mathbf{u}_n(q)\ge \mathbf{u}_{n-|w|}(p)$. Hence we obtain the conclusion by dividing both sides by $P_{q_0}(n)\theta^n$.

We can split the set of states of $\mathcal{M}_L$ into two subsets
$Q_0=\{q\mid a_q=0\}$ and $Q_{>0}=\{q\mid a_q>0\}$. If we consider
only the states of $Q_{>0}$ and the corresponding edges connecting
those states, we obtain a new automaton accepting a new language $L'$.
Representations of real numbers for the numeration system built upon
$L$ or $L'$ are the same except that for the system built on $L'$ a
real number has at most two representations (only when it is the endpoint
of some interval $I_w$).

\begin{Exa}  
  We consider the language $L$ accepted by the automaton depicted in
  Figure \ref{fig:autom4}. This language is such that the number of
  words beginning with $b$ (resp. $a$ or $c$) has a polynomial (resp.
  exponential) behavior.
    \begin{figure}[h!tbp]
    \begin{center}
      \includegraphics{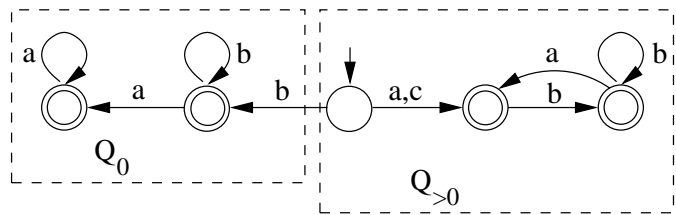}
      \caption{A trim minimal automaton.}
      \label{fig:autom4}
    \end{center}
  \end{figure}
  This means that the length of the interval $I_b$ is zero
  (computations are given in \cite[Example 6]{LR}).  Therefore the
  greatest word in the lexicographical ordering of
  $\mathcal{L}_\infty$ beginning with $a$ represents the same real
  number $x$ as any word in $\mathcal{L}_\infty$ beginning with $b$ or
  the smallest word beginning with $c$.  Removing the states of $Q_0$,
  gives a new language $L'$. In the numeration system built upon $L'$,
  the number $x$ has exactly two representations: the greatest word
  beginning with $a$ and the smallest one beginning with $c$. In this
  latter system, if $w$ is prefix of an infinite number of words in
  $L$ then the length of the interval $I_w$ is strictly positive.
\end{Exa} 

In the following of this paper, we shall assume that $a_q>0$ for all
states $q$ in $\mathcal{M}_L$ except possibly for the sink state.

\section{Determining the numbers having an ultimately periodic representation}\label{sec:sec6}
Let us have a closer look at those intervals $I_w$ (for the details,
the reader is referred to \cite{LR}). If $w\in\Sigma^\ell$ is prefix of
an infinite number of words in $L$ then the interval $I_w$ is given by
\begin{equation}\label{eq:interval}
\biggl[\lim_{n\to\infty}
\biggl(\frac{\mathbf{v}(n-1)}{\mathbf{v}(n)}+\sum_{\substack{m\in\Sigma^\ell\\m<w}}
\frac{\mathbf{u}_{q_0.m}(n-\ell)}{\mathbf{v}(n)}\biggr),\lim_{n\to\infty}
\biggl(\frac{\mathbf{v}(n-1)}{\mathbf{v}(n)}+\sum_{\substack{m\in\Sigma^\ell\\m\le
    w}} \frac{\mathbf{u}_{q_0.m}(n-\ell)}{\mathbf{v}(n)}\biggr)\biggr]
\end{equation}
Using the fact that (see \cite[Proposition 5]{LR})
\begin{equation}
  \label{eq:prop5}
  \left\{\begin{array}{l}
\lim_{n\to\infty} \mathbf{v}_q(n)/\mathbf{v}(n)=a_q\cr
\lim_{n\to\infty} \mathbf{u}_q(n)/\mathbf{v}_q(n)=(\theta-1)/\theta\cr
\end{array}\right.
\end{equation}
the interval $I_w$ 
be rewritten as
\begin{equation}\label{eq:3}
\biggl[\frac{1}{\theta}+ \frac{\theta -1}{\theta^{\ell+1}} \, 
\sum_{\substack{m\in\Sigma^\ell\\m<w}} a_{q_0.m} ,\frac{1}{\theta}+ \frac{\theta -1}{\theta^{\ell+1}} \, 
\sum_{\substack{m\in\Sigma^\ell\\m\le w}} a_{q_0.m}\biggr].
\end{equation}
Notice that this formulation differs slightly from \cite{LR} because
we have here $a_{q_0}=1$ and the others $a_q$'s are strictly positive
(except for the sink).  Observe also that the length of $I_w$ is
$\frac{\theta -1}{\theta^{\ell+1}}\, a_{q_0.w} >0$.

\begin{Rem}
  Notice that if $w\in\Sigma^*$ is prefix of an infinite number of
  words in $L$ then $q_0.w$ is a coaccessible state (so it cannot be
  the sink) and with our assumptions, $a_{q_0.w}>0$.
\end{Rem}

\begin{Exa}\label{exa:cont}
    Consider the numeration system associated to the language accepted
    by the automaton depicted in Figure \ref{fig:autom5}.
   \begin{figure}[h!tbp]
    \begin{center}
      \includegraphics{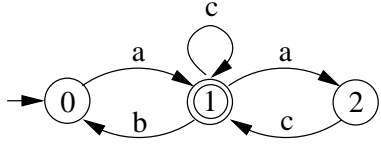}
      \caption{A trim minimal automaton.}
      \label{fig:autom5}
    \end{center}
  \end{figure}
  Here, easy computations show that $\theta=2$, $a_{q_0}=a_{q_2}=1$
  and $a_{q_1}=2$ (to obtain the $a_q$'s, one has only to compute the
  eigenvectors of the eigenvalue $\theta$ of the adjacency matrix).
  Any word in $\mathcal{L}_\infty$ begins with $a$, so $I_a=[1/2,1]$
  (instead of this reasoning, formula \eqref{eq:3} could also be used
  to compute the values of the endpoints of $I_a$). This interval is
  partitioned into three parts,
  $$I_{aa}=[1/2,5/8],\quad I_{ab}=[5/8,3/4],\quad I_{ac}=[3/4,1].$$
  Thus if a real number $x$ belongs to $I_{a\sigma}$ then we have an
  infinite word representing $x$ beginning with $a\sigma$,
  $\sigma\in\Sigma$.  For the next step, we have
$$I_{aac}=I_{aa}=[1/2,5/8],\quad I_{aba}=I_{ab}=[5/8,3/4]$$
and $I_{ac}$ is split into three parts
$$I_{aca}=[3/4,5/6],\quad I_{acb}=[5/6,7/8],\quad I_{aca}=[7/8,1].$$
\end{Exa} 

Actually the form of the partition of an interval $I_w$ into intervals
$I_{w\sigma}$ depends only on the state $q_0.w$ and not on the word
$w$ itself. 
\begin{Def}
  If $0\le \lambda<\mu\le 1$, the strictly increasing
function 
$$f_{[\lambda,\mu]}:[\lambda,\mu]\to[0,1]:x\mapsto
\frac{x-\lambda}{\mu-\lambda}$$
maps the interval $[\lambda,\mu]$ onto
$[0,1]$. If $z$ belongs to $[\lambda,\mu]$ then we say that
$f_{[\lambda,\mu]}(z)$ is the {\it relative position} of $z$ inside
$[\lambda,\mu]$. We denote by $L_w$ (resp. $U_w$) the lower (resp.
upper) bound of the interval $I_w$.
\end{Def}

Roughly speaking, the next proposition states that two intervals
corresponding to the same state are homothetic. But first, we need a
technical lemma.

\begin{Lem}\label{lem:rel} We have, for all states $q$,
  $$\sum_{\sigma\in\Sigma} a_{q.\sigma}=\theta\, a_q$$
in particular, $\sum_{\sigma\in\Sigma} a_{q_0.\sigma}=\theta$.
\end{Lem}

\begin{proof} 
  Clearly, $I_w=\cup_{\sigma\in\Sigma} I_{w\sigma}$ and the length of $I_w$
  is equal to $\sum_{\sigma\in\Sigma} |I_{w\sigma}|$. The conclusion
  follows directly from \eqref{eq:3}.
\end{proof}

\begin{Pro}\label{pro:relative} 
  Let $m$ and $w$ be two words such that $q_0.m=q_0.w$. For all
  $\sigma\in\Sigma$, the interval $I_{m\sigma}$ exists\footnote{An
    interval $I_w$ exists if there exists a word in
    $\mathcal{L}_\infty$ having $w$ as prefix. This means that $w$ is
    prefix of an infinite number of words in $L$.} iff $I_{w\sigma}$
  exists and the relative position of $L_{m\sigma}$ (resp.
  $U_{m\sigma}$) inside $I_m$ is equal to the relative position of
  $L_{w\sigma}$ (resp. $U_{w\sigma}$) inside $I_w$.
\end{Pro}

\begin{proof} 
  The interval $I_{m\sigma}$ exists if $a_{q_0.m\sigma}>0$. Since
  $q_0.m=q_0.w$ then for all $\sigma\in\Sigma$, $a_{q_0.m\sigma}>0$
  iff $a_{q_0.w\sigma}>0$.  Using \eqref{eq:3}, the relative position
  of $L_{m\sigma}$ inside $I_m$ is given by
  $$\biggl(L_{m\sigma}-\frac{1}{\theta}- \frac{\theta
    -1}{\theta^{|m|+1}} \, \sum_{\substack{u\in\Sigma^{|m|}\\u<m}}
  a_{q_0.u}\biggr) / \biggl(\frac{\theta -1}{\theta^{|m|+1}} \,
    a_{q_0.m}\biggr) $$
  this can be rewritten as
  $$\frac{\theta^{|m|}}{(\theta-1)\, a_{q_0.m}} (\theta\, 
  L_{m\sigma}-1)-\sum_{\substack{u\in\Sigma^{|m|}\\u<m}}
  \frac{a_{q_0.u}}{a_{q_0.m}}$$
  and using the definition of $L_{m\sigma}$, 
  we get 
\begin{equation}\label{eq:compare1}
\frac{1}{a_{q_0.m}}\biggl(\frac{1}{\theta} \sum_{\substack{u\in\Sigma^{|m|+1}\\u<m\sigma}}
a_{q_0.u}-\sum_{\substack{u\in\Sigma^{|m|}\\u<m}}
a_{q_0.u}\biggr).
\end{equation}
First notice that the sum over the words $u$ of length $|m|+1$ and
lexicographically less than $m\sigma$ can be split into two subsets:
the words $m\tau$ with $\tau<\sigma$ and the words having a prefix of
length $|m|$ lexicographically less than $m$. So \eqref{eq:compare1}
can be written
$$\frac{1}{a_{q_0.m}}\biggl[\frac{1}{\theta}
  \biggl(\sum_{\substack{\tau\in\Sigma\\ \tau<\sigma}}
    a_{q_0.m\tau}+\sum_{\substack{u\in\Sigma^{|m|+1}\\pref_{|m|}(u)<m}}
    a_{q_0.u}\biggr)-\sum_{\substack{u\in\Sigma^{|m|}\\u<m}}
  a_{q_0.u}\biggr]$$
where $pref_{|m|}(u)<m$ means that the prefix of length
$|m|$ of $u$ is lexicographically less than $m$. To conclude the
proof, notice that
$$
\sum_{\substack{u\in\Sigma^{|m|+1}\\pref_{|m|}(u)<m}}
a_{q_0.u}=\sum_{\substack{u'\in\Sigma^{|m|}\\u'<m}}\sum_{\sigma\in\Sigma}
a_{q_0.u'\sigma}$$
and using Lemma \ref{lem:rel}, we have
$$\frac{1}{\theta}
\sum_{\substack{u\in\Sigma^{|m|+1}\\pref_{|m|}(u)<m}}
a_{q_0.u}-\sum_{\substack{u\in\Sigma^{|m|}\\u<m}}
  a_{q_0.u}=\frac{1}{\theta}\sum_{\substack{u'\in\Sigma^{|m|}\\u'<m}}\theta 
a_{q_0.u'} -\sum_{\substack{u\in\Sigma^{|m|}\\u<m}} a_{q_0.u} =0.$$
So the relative position of $L_{m\sigma}$ inside $I_m$ is 
\begin{equation}
\frac{1}{\theta\, a_{q_0.m}}
  \sum_{\substack{\tau\in\Sigma\\ \tau<\sigma}}
    a_{q_0.m\tau}
\end{equation}
and depends only on the state $q_0.m$.
\end{proof}

\begin{Exa}
  Continuing Example \ref{exa:cont}. We have $q_0.a=q_0.ac=q_1$.
  Observe that $I_a$ is split into three parts ($I_{aa}$, $I_{ab}$ and
  $I_{ac}$) and the relative positions of $5/8$ and $3/4$ inside $I_a$
  are respectively $1/4$ and $1/2$. In the same way, $I_{ac}$ is split
  into three parts ($I_{aca}$, $I_{acb}$ and $I_{acc}$) the relative
  positions of $5/6$ and $7/8$ inside $I_{ac}$ are also respectively
  $1/4$ and $1/2$.
\end{Exa}

\begin{The}\label{the:per} A real number $x$ has an ultimately periodic representation if and only if there exist two words $m$ and $w$ such that
\begin{enumerate}
\item $m$ is a prefix of $w$,
\item $x$ belongs to $I_w\subset I_m$,
\item $q_0.m=q_0.w$,
\item the relative position of $x$ inside $I_m$ is equal to the relative position of $x$ inside $I_w$.
\end{enumerate}  
\end{The}

\begin{proof}
  The condition is sufficient. If $x$ belongs to $I_m$ then a
  representation of $x$ has $m$ has prefix. The following letter of
  the representation depends only on the relative position of $x$
  inside $I_m$. Assume that this letter is $\sigma$ (i.e., $x \in
  I_{m\sigma}$). Thanks to Proposition \ref{pro:relative}, since
  $q_0.m=q_0.w$ and the relative position of $x$ inside $I_m$ is equal
  to the relative position of $x$ inside $I_w$, it is clear that $x$
  belongs to $I_{w\sigma}$. The same arguments can be used with
  $I_{m\sigma}$ and $I_{w\sigma}$ and so on. Actually, if $w=mv$ then
  the representation of $x$ is $mv^\omega$.
  
  Assume now that $x$ has an ultimately periodic representation
  $w=w_0w_1\cdots=uv^\omega $. It is well known that $I_{w_0\cdots
    w_j}\subset I_{w_0\cdots w_i}$ for $i<j$. Since the automaton
  $\mathcal{M}_L$ is finite, there exist infinitely many indices
  $i_1<i_2<\ldots$ and a constant $C$ such that $$q_0.w_0\cdots
  w_{i_1}=q_0.w_0\cdots w_{i_2}=\ldots,\quad i_1\ge |u| \quad
  \text{and}\quad \forall k\ge 1,\ i_{k+1}-i_{k}=C|v|.$$
  Assume that
  for any pair $i_j<i_k$ of such indices the relative position of $x$
  inside $I_{w_0\cdots w_{i_j}}$ is different from its relative
  position inside $I_{w_0\cdots w_{i_k}}$. To conclude the proof, we
  have to show that the representation of $x$ is not ultimately
  periodic. If $x$ belongs to some $I_W$ ($W=w_0\cdots w_{i_j}$), then
  $x$ has a representation beginning with $W$ and to determine the
  following letter of this representation, the interval $I_W$ is
  divided into intervals $I_{W\sigma}$. This process of dividing
  intervals into smaller intervals is repeated continuously and we
  already know that the length of $I_W$ is $\frac{\theta
    -1}{\theta^{|W|+1}} a_{q_0.W}\in \mathcal{O}(\theta^{-|W|})$.
  Since relative positions of $x$ inside $I_{w_0\cdots w_{i_j}}$ and
  $I_{w_0\cdots w_{i_k}}$ are different, by Proposition
  \ref{pro:relative} there exist $n$ and $\sigma\neq\tau$ such that
  $$x\in I_{w_0\cdots w_{i_j}w_{i_j+1}\cdots w_{i_j+n}\sigma}\quad
  {\text and }\quad x\in I_{w_0\cdots w_{i_k}w_{i_k+1}\cdots
    w_{i_k+n}\tau}.$$
  Therefore $w$ is not ultimately periodic.
\end{proof}

Since the form of the intervals $I_w$ depends only on the states of $\mathcal{M}_L$, we can define some dynamical system.

\begin{Def}\label{def:dyn}
  For each $q\in Q$, we define a partition of $A_q=[0,1]$ into
  intervals $A'_{q,\sigma}$ in the following way. Since
  $\mathcal{M}_L$ is accessible, there exists $w$ such that $q=q_0.w$.
  For each $\sigma\in\Sigma$ such that $I_{w\sigma}$ exists consider
  the relative position $\ell_{q,\sigma}$ (resp. $u_{q,\sigma}$) of
  $L_{w\sigma}$ (resp. $U_{w\sigma}$) inside $I_w$. We denote
$$A'_{q,\sigma}=[\ell_{q,\sigma},u_{q,\sigma}).$$
(If $\sigma$ is the
largest letter such that $I_{w\sigma}$ exists then
$A'_{q,\sigma}=[\ell_{q,\sigma},u_{q,\sigma}]=[\ell_{q,\sigma},1]$.)

Let us define a
function $h:Q\times [0,1] \to Q\times [0,1] : (q,x) \mapsto (q',x')$ in the following manner. Since we have a partition of $A_q$, there exists a unique letter $\sigma$ such that $x\in A'_{q,\sigma}$ and therefore
$$\left\{\begin{array}{l}
q'=q.\sigma\cr
x'=f_{A'_{q,\sigma}}(x)\cr
\end{array}\right.$$
where $f_{A'_{q,\sigma}}(x)$ denotes the relative position of $x$
inside $A'_{q,\sigma}$. (For the interested reader, this dynamical
system looks like up to some extend to interval exchange
transformations \cite{iex1,iex2}.)
\end{Def}

\begin{Exa}
    Continuing Example \ref{exa:cont}. The interval $A_{q_0}=[0,1]$ is
    partitioned into a single interval $A'_{q_0,a}=[0,1]$. The
    interval $A_{q_1}=[0,1]$ is partitioned into
    $A'_{q_1,a}=[0,1/4)$, $A'_{q_1,b}=[1/4,1/2)$ and
    $A'_{q_1,c}=[1/2,1]$. Finally $A_{q_2}=[0,1]$ is partitioned
    into a single interval $A'_{q_2,c}=[0,1]$.
\end{Exa}

Let us now present two equivalent algorithms for computing the
representation of a real number. We recall that
$I_\varepsilon=[1/\theta,1]$. We denote by $f_I(x)$ the relative
position of $x$ inside the interval $I$.

\begin{Alg}\label{algo:1}
  Let $x\in [1/\theta,1]$\\
  {\tt \indent {\bf Initialization}\\
    \indent \indent $q\leftarrow q_0$\\
    \indent \indent $w\leftarrow \varepsilon$\\
    \indent \indent $y\leftarrow f_{I_w}(x)$\\
    \indent {\bf repeat}\\
    \indent \indent Determine the letter $\sigma\in\Sigma$ such that $y\in A'_{q,\sigma}$.\\
    \indent \indent $q\leftarrow q.\sigma$\\
    \indent \indent $w\leftarrow$ concat$(w,\sigma)$\\
    \indent \indent $y\leftarrow f_{I_w}(x)$\\
    \indent {\bf until a stop condition}}.
\end{Alg} 

The stop condition of the algorithm can be a fixed number of $k$
iterations to determine the first $k$ letters of a representation. One
could check if a representation is ultimately periodic. Indeed, if
we denote by $q_n$ and $y_n$ the values of the variables $q$ and $y$
during the $n$th iteration of the algorithm then thanks to Theorem
\ref{the:per}, a representation is ultimately periodic if there exist
$i\neq j$ such that $q_i=q_j$ and $y_i=y_j$. A variant of this
algorithm is the following one.

 \begin{Alg}\label{algo:2}
     Let $x\in [1/\theta,1]$\\
     {\tt \indent{\bf Initialization}\\
       \indent\indent $q\leftarrow q_0$\\
       \indent\indent $w\leftarrow \varepsilon$\\
       \indent\indent $I\leftarrow [1/\theta,1]$\\
       \indent\indent $x\leftarrow f_I(x)$\\
       \indent{\bf repeat}\\
       \indent\indent Determine the letter $\sigma\in\Sigma$ such that $x\in A'_{q,\sigma}$.\\
       \indent\indent $q\leftarrow q.\sigma$\\
       \indent\indent $w\leftarrow$ concat$(w,\sigma)$\\
       \indent\indent $I\leftarrow A'_{q,\sigma}$\\
       \indent\indent $x\leftarrow f_I(x)$\\
       \indent{\bf until a stop condition}}.  
\end{Alg} 

In this latter algorithm, a periodicity in the representation is found
when $q_i=q_j$ and $x_i=x_j$.

\begin{Exa} 
  Continuing Example \ref{exa:cont}. We can try to obtain the representation of $x=4/7$ using Algorithm \ref{algo:1}. The computations are given in Table \ref{tab:1}. At each step of the procedure, the intervals $I_w$ are given below their corresponding $A'_{q,\sigma}$. Observe that at two steps $i<j$ of the algorithm, if we have the same state $q$ then the intervals $A'_{q,\sigma}$ are the same but the intervals $I_w$ are getting smaller. 
\begin{table}[h!tbp]
$$\begin{array}{|c|c|c||c|c|c||l|}
\hline
 q & w & f_{I_w}(x) & A'_{q,a} & A'_{q,b} & A'_{q,c} & \cr
  &  &  & I_{wa} & I_{wb} & I_{wc} & \cr
\hline
 0 & \varepsilon & 1/7 & [0,1] & --- & --- &  \cr
 & &     & [1/2,1] & --- & --- &  \cr 
\hline
  1  & a  & 1/7 & [0,1/4) & [1/4,1/2) & [1/2,1] & 1/7 < 1/4  \cr 
  &  &     & [1/2,5/8] & [5/8,3/4] & [3/4,1] & 4/7 <5/8 \cr
\hline
 2 & aa & 4/7 & --- & --- & [0,1) &  \cr
 & & & --- & --- & [1/2,5/8] &  \cr
\hline
 1 & aac & 4/7 & [0,1/4) & [1/4,1/2) & [1/2,1] & 1/2 < 4/7 \cr
& & & [1/2,17/32] & [17/32,9/16] & [9/16,5/8]& 9/16 < 4/7 \cr
\hline
 1 & aacc & 1/7 & & & & \cr
\hline
\end{array}$$
\caption{Representation of $x=4/7$.}\label{tab:1}
\end{table}
Thanks to Theorem \ref{the:per}, $4/7$ is represented by
$a(acc)^\omega$.  The iteration of the function $h$ introduced in
Definition \ref{def:dyn} can also be used to find the real numbers
having an ultimately periodic representation, indeed here we have
 $$(q_0,1/7)\stackrel{h}{\mapsto}\underline{(q_1,1/7)}
\stackrel{h}{\mapsto}(q_2,4/7)
 \stackrel{h}{\mapsto}(q_1,4/7)\stackrel{h}{\mapsto}\underline{(q_1,1/7)}.$$
\end{Exa}

\section{A characterization}\label{sec:sec7}
With the abstraction of the previous section, we can summarize the
informations needed to compute representations of real numbers: a
finite number of partitions $A_q=\cup A'_{q,\sigma}$ of the interval
$[0,1]$ and the transition function of $\mathcal{M}_L$.  The aim of
this section is to give a characterization of the real numbers having
an ultimately periodic representation in terms of composition of some
affine functions.

\begin{Def}
  Let us consider the automaton $\mathcal{F}_L$ defined as follows
\begin{itemize}
\item The set of states is $\{A_q\mid q\in Q\}$.
\item If $A_q$ is partitioned into $A'_{q,\sigma_1}\cup \ldots \cup
  A'_{q,\sigma_t}$ then we have an edge labeled by
  $f_{A'_{q,\sigma_i}}$ from $A_q$ to $A_{q.\sigma_i}$ (where the dot
  in $q.\sigma_i$ denotes the transition function of $\mathcal{M}_L$)
  for $i=1,\ldots ,t$.
\item All the states are final.
\end{itemize}
Except that the initial state is not important in what follows and
that the labels of the edges have changed, $\mathcal{F}_L$ is more or
less a copy of $\mathcal{M}_L$.
\end{Def}

\begin{Exa} 
    Continuing Example \ref{exa:cont}. Here,
    $f_{A'_{q_0,a}}=f_{A'_{q_2,c}}=id$ and
 $$\left\{\begin{array}{lrrl}
   f_{A'_{q_1,a}}:&[0,1/4]\to [0,1]&:x\mapsto &4x, \cr
   f_{A'_{q_1,b}}:&[1/4,1/2]\to [0,1]&:x\mapsto &4x-1,\cr
   f_{A'_{q_1,c}}:&[1/2,1]\to [0,1]&:x\mapsto &2x-1.\cr
\end{array}\right.$$
The automaton $\mathcal{F}_L$ is depicted in Figure \ref{fig:autom6}.
For the sake of simplicity, $f_{A'_{q_1,\sigma}}$ is denoted by
$f_\sigma$, for $\sigma=a,b,c$ (since it does not lead to any
confusion). We also put an index $a$ or $c$ to $id$ to remember the
corresponding letter.
\begin{figure}[h!tbp]
    \begin{center}
      \includegraphics{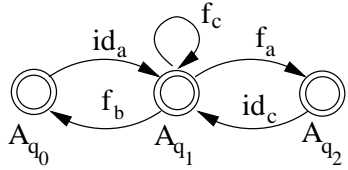}
      \caption{The automaton $\mathcal{F}_L$.}
      \label{fig:autom6}
    \end{center}
  \end{figure}
  A path $f_1\cdots f_t$ in $\mathcal{F}_L$ corresponds to the
  composition of affine functions $f_t\circ \cdots \circ f_1$ in
  reversed order. Through $\mathcal{F}_L$, we can determine the real
  numbers having ultimately periodic representations. 
  
  Indeed, if we consider a cycle $f_1\cdots f_t$ in $\mathcal{F}_L$
  starting in $A_{q_0}$ then, in view of Algorithm \ref{algo:2}, if
  the unique fixed point of the corresponding function $F= f_t\circ
  \cdots \circ f_1$ is $x$ then $f_I^{-1}(x)$, with $f_I:x\mapsto
  \frac{\theta x-1}{\theta-1}$ and
  $f_I^{-1}(y)=\frac{(\theta-1)y+1}{\theta}$, has an ultimately
  periodic representation (we are back in the initial state $q_0$ and
  since $x=F(x)$, we have the same initial value; due to the 
  initialization step in Algorithm \ref{algo:2}, we have to apply
  $f_I^{-1}$ once). As an example, the fixed points of $f_b\circ id_a$,
  $f_b\circ id_c \circ f_a\circ id_a$ and $f_b\circ f_c \circ id_c \circ f_a
  \circ id_a$ are respectively $1/3$, $1/15$ and $5/31$ and therefore
  $2/3$, $8/15$ and $18/31$ have ultimately periodic representations.
  From the path in $\mathcal{F}_L$, we also know these
  representations: $(ab)^\omega$, $(aacb)^\omega$ and
  $(aaccb)^\omega$.
  
  We can also consider a cycle $f_1\cdots f_t$ in $\mathcal{F}_L$
  starting in $A_{q_1}$ instead of $A_{q_0}$ and a path $g_1\cdots
  g_s$ from $A_{q_0}$ to $A_{q_1}$. Once again, let $x$ be the fixed
  point of $F= f_t\circ \cdots \circ f_1$. From Algorithm
  \ref{algo:2}, the number $f_I^{-1}\circ g_1^{-1}\circ \cdots \circ
  g_s^{-1}(x)$ has an ultimately periodic representation. As an
  example, consider $F=f_c \circ id_c\circ f_a \circ f_c$ having $3/5$
  has fixed point. A trivial path from $A_{q_0}$ to $A_{q_1}$ is given
  by $id_a$, so $f_I^{-1}(3/5)=4/5$ has an ultimately periodic
  representation: $a(cacc)^\omega$. Another path in $\mathcal{F}_L$
  from $A_{q_0}$ to $A_{q_1}$ is $(id_a,f_a,id_c)$, so $f_I^{-1}\circ
  f_a^{-1}(3/5)=23/40$ is represented by $aac(cacc)^\omega$.
\end{Exa}

Let us introduce some notation. Let 
$$\phi_q=\{w \mid \delta_{\mathcal{F}}(A_q,w)=A_q\},$$
where $\delta_{\mathcal{F}}$ is the transition function of $\mathcal{F}_L$.
If $w=f_1\cdots f_t$ belongs to $\phi_q$, we denote by $F_{q,w}$ the composed
function (in reversed order) $f_t\circ \cdots \circ f_1$ corresponding to $w$. 
Let
$$\nu_q=\{w \mid \delta_{\mathcal{F}}(A_{q_0},w)=A_q\},$$
If
$w=f_1\cdots f_s$ belongs to $\nu_q$, we denote by $(F_{q,w})^{-1}$ the
composition of the inverse functions 
$f_1^{-1}\circ \cdots \circ f_s^{-1}$ corresponding to $w$.

\begin{The} 
  Let $L$ be a regular language satisfying our basic assumptions. Set
  $$f_I^{-1}:y\mapsto\frac{(\theta-1)\, y+1}{\theta}.$$
  The set of
  real numbers having an ultimately periodic representation is given
  by
$$\{ f_I^{-1}\circ (F_{q,w})^{-1} (x) \mid \exists q\in Q, z\in  \phi_q, w\in\nu_q :   x=F_{q,z}(x)\}.$$
\end{The}

\begin{proof}
  This is a direct consequence of Algorithm \ref{algo:2}.
\end{proof}

\begin{Rem}
    For all states $q$, the languages $\phi_q$ and $\nu_q$ over a
    finite alphabet of functions are regular.
\end{Rem}

\section{Equivalence with $\theta$-development}\label{sec:sec8}
Let $\theta>1$ be a Pisot number. To this number corresponds a unique
positional and linear Bertrand number system $U_\theta=(U_n)_{n\in
  \mathbb{N}}$ having its characteristic polynomial equal to the
minimal polynomial of $\theta$ \cite{BH}.  We denote by $L$ the
language $\rho_U(\mathbb{N})$ of all the normalized representations
computed by the greedy algorithm (without leading zeroes) \cite{Fr}.
In this section we show that this latter language $L$ satisfies the
hypotheses given in Section \ref{sec:2}. We also prove that the
representations of real numbers in the abstract numeration system
built upon $L$ and the classical $\theta$-developments of the numbers
in $[\frac{1}{\theta},1]$ coincide. (For a presentation of the
$\theta$-development, we refer the reader to \cite[Chapter 7]{Lot} or
\cite{Pa}.)

\begin{Def}
Recall that a positional numeration system $U=(U_n)_{n\in \mathbb{N}}$
is said to be a {\it Bertrand numeration system} if
$$\forall n\in \mathbb{N}, w\, 0^n \in \rho_U(\mathbb{N})
\Leftrightarrow w\in \rho_U(\mathbb{N}).$$
As an example, the $k$-ary
number system is a Bertrand system.
\end{Def}

\begin{Exa} 
  The golden ration $\tau=\frac{1+\sqrt{5}}{2}$ is a Pisot number,
  indeed its minimal polynomial is $P(X)=X^2-X-1$ and the other root
  of $P$ has modulus less than one. The polynomial $P$ is also the
  characteristic polynomial of the linear recurrence relation defined
  by
  $$U_{n+2}=U_{n+1}+U_n,\quad n\in\mathbb{N}.$$
  If we consider the
  initial conditions $U_0=1$ and $U_1=2$, then as a consequence of the
  greedy algorithm, the set of representations of the integers is
  $\rho_U(\mathbb{N})=\{\varepsilon\}\cup 1\{0,01\}^*$. Due to the
  particular form of the language $\rho_U(\mathbb{N})$, it is clear
  that this system (namely the Fibonacci system) is the linear
  Bertrand number system associated to $\tau$.
\end{Exa}

Consider an arbitrary Pisot number $\theta$. It is well known that the
$\theta$-development of one is finite or ultimately periodic
\cite{KS}. In the first case,
$e_\theta(1)=t_1\cdots t_m$ 
and we define, as usual,
$$e_\theta^*(1)=(t_1\cdots t_{m-1}(t_m-1))^\omega.$$
It is clear that we still have 
$$1=\frac{t_1}{\theta}+\frac{t_2}{\theta^2}+\cdots +\frac{t_m-1}{\theta^m}+\frac{t_1}{\theta^{m+1}}+ \cdots.$$

Let $\beta>1$ be a real number. The set $D_\beta$ of
$\beta$-developments of numbers in $[0,1)$ is characterized as
follows.
\begin{The}\label{parry}{\rm \cite{Pa}} Let $\beta>1$ be a real number. 
A sequence $(x_n)_{n\ge 1}$ belongs to $D_\beta$ if and only if for all $i\in\mathbb{N}$, the shifted sequence $(x_{n+i})_{n\ge 1}$ is lexicographically less than the sequence $e_\beta(1)$ or $e_\beta^*(1)$ whenever  
$e_\beta(1)$ is finite.
\end{The}

For any real number $\beta>1$, we denote by $F(D_\beta)$, the set of finite factors of the sequences in $D_\beta$.  Bertrand numeration systems are characterized by  the theorem of Bertrand given below. 
Notice that $U$ is not necessarily linear. 
\begin{The}{\rm \cite{Ber}} Let $U=(U_n)_{n\in\mathbb{N}}$ be a positional numeration system. Then $U$ is a Bertrand numeration system if and only if there exists a real number $\beta >1$ such that $0^*\rho_U(\mathbb{N})=F(D_\beta)$. In this case,
if $e_\beta(1)=(d_n)_{n\ge 1}$ (or $e_\beta^*(1)=(d_n)_{n\ge 1}$ whenever 
$e_\beta(1)$ is finite) then $U_0=1$ and
$$U_n=d_1\, U_{n-1} + d_2\, U_{n-2}+\cdots +d_n \, U_0+1,\ n\ge 1.$$
\end{The}

Let $\theta >1$ be a Pisot number. First we assume that $e_\theta(1)$
is ultimately periodic; there exist minimal integers $N\ge 0$, $p\ge 1$
such that
$$e_\theta(1)=t_1\cdots t_N\, (t_{N+1}\cdots t_{N+p})^\omega.$$
 The Bertrand numeration system $U_\theta=(U_n)_{n\in\mathbb{N}}$ belonging to the class of positional systems related to $\theta$ is a linear numeration system satisfying the recurrence relation
\begin{eqnarray*}
U_n&=& t_1\, U_{n-1}+\cdots + t_{p-1}\, U_{n-p+1} +(t_p+1)\, U_{n-p} \\
 & & +(t_{p+1}-t_1)\, U_{n-p-1}+\cdots +(t_{N+p}-t_N)\, U_{n-N-p},\ n\ge N+p.
\end{eqnarray*}
(In other words, $(U_n)_{n\in\mathbb{N}}$ satisfies the canonical beta
polynomial of $\theta$ \cite{Ho}.) In what follows, $\theta$ is given and we
denote $U_\theta$ simply by $U$.  

\begin{figure}[h!tbp]
\begin{center}
\includegraphics{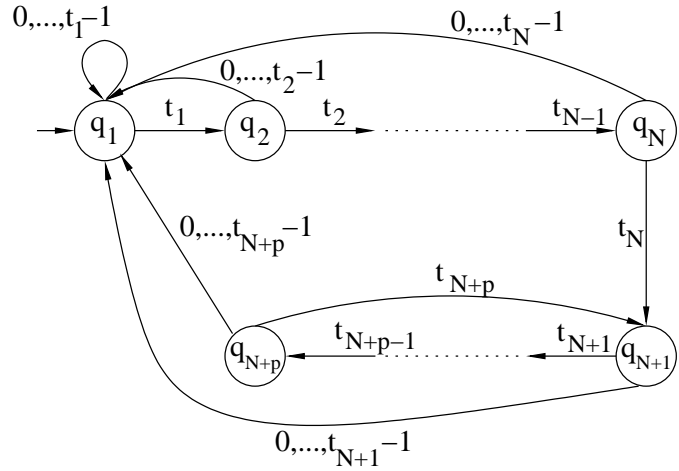}
\caption{Automaton recognizing $F(D_\theta)=0^*\rho_U(\mathbb{N})$.}\label{fig:dtheta}
\end{center}
\end{figure}

The main point is the following. Since $\theta$ is a Pisot number, the
set $F(D_\theta)=0^*\rho_U(\mathbb{N})$ is recognizable by a finite
automaton ${\mathcal A}$ \cite{FS} (i.e., the $\theta$-shift is
sofic). This automaton has $N+p$ states $q_1,\ldots ,q_{N+p}$. For
each $i\in \{1,\ldots ,N+p\}$, there are edges labeled by $0,1,\ldots
,t_i-1$ from $q_i$ to $q_1$, and an edge labeled $t_i$ from $q_i$ to
$q_{i+1}$ if $i<N+p$. Finally, there is an edge labeled $t_{N+p}$ from
$q_{N+p}$ to $q_{N+1}$. All states are final and $q_1$ is the initial
state. The set $F(D_\theta)$ is recognized by the automaton depicted
in Figure \ref{fig:dtheta} (the sink is not represented).

\begin{figure}[h!tbp]
\begin{center}
\includegraphics{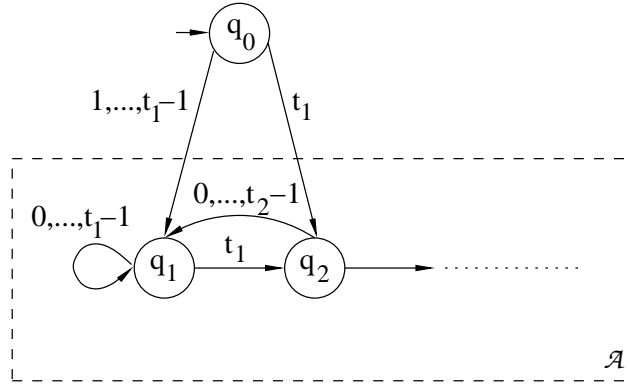}
\caption{The automaton $\mathcal{A}'$ 
recognizing $\rho_U(\mathbb{N})$.}\label{fig:dtheta2}
\end{center}
\end{figure}

In an abstract numeration system, allowing leading zeroes changes the
representations (indeed, $0w$ is genealogically greater than $w$ then
$\val{S}(0w)>\val{S}(w)$, see for instance \cite[Example 1]{LR2}).
Therefore, we modify slightly the automaton ${\mathcal A}$ to obtain
an automaton ${\mathcal A}'$ recognizing exactly $\rho_U(\mathbb{N})$
(i.e., without leading zeroes). To that end, we add a new state $q_0$.
There are edges labeled by $1,\ldots ,t_1-1$ from $q_0$ to $q_1$ and
an edge labeled $t_1$ from $q_0$ to $q_2$.  This state $q_0$ is the
initial state of ${\mathcal A}'$ and is also final. The automaton
${\mathcal A}'$ is sketched in Figure \ref{fig:dtheta2}.

So we consider the abstract system built upon $L=\rho_U(\mathbb{N})$.
To be able to compute the intervals $I_w$ related to this system with
formula \eqref{eq:interval}, our task is now to determine the
different sequences ${\mathbf u}_{q_i}(n)$.  To that end, we use the
specific form of ${\mathcal A}'$. The first word of length $n+1$ in
$\rho_U(\mathbb{N})$ is $1(0)^n$ and its numerical value is $U_n$. In
the same way, $1(0)^{n-1}$ is the first word of length $n$. Therefore,
\begin{equation}\label{eq:uq0}
{\mathbf u}_{q_0}(n)=U_n-U_{n-1}.  
\end{equation}
Since $q_0$ is the initial state of
$\mathcal{A}'$, as usual we write ${\mathbf u}(n)$ and ${\mathbf
  v}(n)$ instead of ${\mathbf u}_{q_0}(n)$ and ${\mathbf v}_{q_0}(n)$.

Since
$\theta$ is a Pisot number and the characteristic polynomial of $U$ is
the minimal polynomial of $\theta$, there exists a real number
$\gamma$ such that
$$U_n\sim \gamma \, \theta^n.$$
For $n\ge 1$, from the form of $\mathcal{A}'$ we deduce that
$${\mathbf u}(n)=(t_1-1)\, {\mathbf u}_{q_1}(n-1)+{\mathbf u}_{q_2}(n-1)$$
and 
\begin{equation}\label{eq:calculu}
{\mathbf u}_{q_1}(n)=t_1\, {\mathbf u}_{q_1}(n-1)+{\mathbf u}_{q_2}(n-1)={\mathbf u}(n)+{\mathbf u}_{q_1}(n-1).
\end{equation}
As a consequence of \eqref{eq:calculu}, since all the states are final
$\mathbf{u}_{q_1}(0)=\mathbf{u}(0)=1$, we find
$${\mathbf u}_{q_1}(n)=\sum_{i=1}^n {\mathbf u}(i)+{\mathbf
  u}_{q_1}(0)={\mathbf v}(n).$$
From \eqref{eq:calculu} we also have
${\mathbf u}_{q_2}(n-1)={\mathbf u}_{q_1}(n)-t_1\, {\mathbf
  u}_{q_1}(n-1)$ and thus ${\mathbf u}_{q_2}(n)={\mathbf v}(n+1)-t_1\,
{\mathbf v}(n)$.  But considering the path in $\mathcal{A}'$, we get
${\mathbf u}_{q_2}(n)=t_2\, {\mathbf u}_{q_1}(n-1)+{\mathbf
  u}_{q_3}(n-1)$. So we find ${\mathbf u}_{q_3}(n)={\mathbf
  v}(n+2)-t_1\, {\mathbf v}(n+1)-t_2\, {\mathbf v}(n)$.  Continuing
this way, for $i\le N+p$
$${\mathbf u}_{q_i}(n)={\mathbf v}(n+i-1)-t_1\, {\mathbf
  v}(n+i-2)-t_2\, {\mathbf v}(n+i-3)-\cdots -t_{i-1} {\mathbf v}(n)
.$$

We are now able to determine the endpoints of the intervals $I_w$. It
is clear from \eqref{eq:uq0} that ${\mathbf v}(n)=U_n \sim \gamma\,
\theta^{n}$.  Since ${\mathbf v}(n-1)={\mathbf v}(n)-{\mathbf u}(n)$
using \eqref{eq:prop5}, it is clear that ${\mathbf v}(n-1)/{\mathbf
  v}(n)\to1/\theta$ if $n\to\infty$ and therefore, for $i\in
\mathbb{N}$
$$\lim_{n\to \infty} \frac{{\mathbf u}_{q_1}(n-i)}{{\mathbf v}(n)}=
\lim_{n\to \infty} \frac{{\mathbf v}(n-i)}{{\mathbf
    v}(n-i+1)}\frac{{\mathbf v}(n-i+1)}{{\mathbf v}(n-i+2)}\cdots
\frac{{\mathbf v}(n-1)}{{\mathbf v}(n)}=\theta^{-i}.$$
In the same
manner,
$$\lim_{n\to \infty} \frac{{\mathbf u}_{q_2}(n-i)}{{\mathbf
    v}(n)}=\theta^{1-i}-t_1\, \theta^{-i}.$$
Continuing this way, for
$j\le N+p$ and $i\in \mathbb{N}$
$$\lim_{n\to \infty} \frac{{\mathbf u}_{q_j}(n-i)}{{\mathbf
    v}(n)}=\theta^{j-i-1}-t_1\, \theta^{j-i-2}- \cdots -t_{j-1}
\theta^{-i}.$$
We can now compute the different intervals $I_w$. The
first words in $\rho_U(\mathbb{N})$ are $$1,\ldots,(t_1-1),t_1,\ 
10,\ldots ,1t_1,20,\ldots , (t_1-1)t_1, t_10,\ldots ,t_1t_2,\ 
100\ldots $$
Using \eqref{eq:interval} we have the intervals
corresponding to words of length one
$$I_j=[\frac{j}{\theta},\frac{j}{\theta}+\lim_{n\to\infty}\frac{{\mathbf
    u}_{q_1}(n-1)}{{\mathbf
    v}(n)}]=[\frac{j}{\theta},\frac{j+1}{\theta}], \ 1\le j<t_1$$
and
$$I_{t_1}=[\frac{t_1}{\theta},\frac{t_1}{\theta}+\lim_{n\to\infty}\frac{{\mathbf u}_{q_2}(n-1)}{{\mathbf v}(n)}]=[\frac{t_1}{\theta},1].$$
For the words of length two, if $1\le j<t_1$ and $0\le k<t_1$ then
$$I_{jk}=[\frac{j}{\theta}+\frac{k}{\theta^2},\frac{j}{\theta}+\frac{k}{\theta^2}+\lim_{n\to\infty}\frac{{\mathbf
    u}_{q_1}(n-2)}{{\mathbf
    v}(n)}]=[\frac{j}{\theta}+\frac{k}{\theta^2},\frac{j}{\theta}+\frac{k+1}{\theta^2}]$$
and
$$I_{j{t_1}}=[\frac{j}{\theta}+\frac{t_1}{\theta^2},\frac{j}{\theta}+\frac{t_1}{\theta^2}+\lim_{n\to\infty}\frac{{\mathbf
    u}_{q_2}(n-2)}{{\mathbf v}(n)}]=
[\frac{j}{\theta}+\frac{t_1}{\theta^2},\frac{j+1}{\theta}].$$
For the
words of length two beginning with $t_1$, we have
$$I_{{t_1}j}=[\frac{t_1}{\theta}+\frac{j}{\theta^2}, \frac{t_1}{\theta}+\frac{j+1}{\theta^2}],\ 0\le j<t_2$$
and 
$$I_{{t_1}{t_2}}=[\frac{t_1}{\theta}+\frac{t_2}{\theta^2},
\frac{t_1}{\theta}+\frac{t_2}{\theta^2}+\lim_{n\to\infty}\frac{{\mathbf
    u}_{q_3}(n-2)}{{\mathbf
    v}(n)}]=[\frac{t_1}{\theta}+\frac{t_2}{\theta^2},1].$$
Continuing
this way, it is straightforward computation to see that we have three
situations: 
\begin{enumerate}
\item if $w=w_0\cdots w_r$ with $q_i=q_0.w_0\cdots w_{r-1}$ and
$w_r< t_i$ then
$$I_w=[\sum_{i=0}^r w_i \theta^{-i-1},\sum_{i=0}^r w_i
\theta^{-i-1}+\theta^{-r-1}]$$

\item if $w=w_0\cdots w_r w_{r+1}\cdots w_{r+s}$ is such that
  $q_i=q_0.w_0\cdots w_{r-1}$, $w_r< t_i$ and $w_{r+1}\cdots w_{r+s}$
  is the maximal word read from $q_0.w_0\cdots w_r$ in $\mathcal{A}'$
  (in other words, $q_0.w_0\cdots w_{r}=q_1$ and $w_{r+1}\cdots
  w_{r+s}$ is the prefix of length $s$ of $e_\theta(1)$) then
$$I_w=[\sum_{i=0}^{r+s} w_i \theta^{-i-1},\sum_{i=0}^r w_i
\theta^{-i-1}+\theta^{-r-1}].$$

\item finally, if $w=w_0\cdots w_r$ is a prefix of $e_\theta(1)$ then 
$$I_w=[\sum_{i=0}^r w_i \theta^{-i-1},1].$$
\end{enumerate}

Now instead of considering the abstract numeration system built upon
$L=\rho_U(\mathbb{N})$, we can consider the classical
$\theta$-development of a real number $x\in[1/\theta,1]$. The first
digit of $e_\theta(x)$ is an integer $j$ belonging to
$\{1,2,\ldots,t_1\}$. Since $\theta$-developments are computed through
the greedy algorithm, it is clear that the first digit is $j<t_1$ if
and only if $x\in I'_j=[j/\theta,(j+1)/\theta)$ and it is $t_1$ if and
only if $x\in I'_{t_1}=[t_1/\theta,1]$. So the interval $I_j$ for the
abstract numeration systems considered above and the intervals $I'_j$
corresponding to the greedy algorithm are the same for the first step
(except that in the abstract system, a real number can have two
representations but we can avoid this ambiguity by considering
intervals of the form $[a,b)$ and therefore the two intervals $I_j$
and $I'_j$ will coincide exactly). By application of the greedy
algorithm, we can compute intervals $I'_w$ such that $x$ belongs to
$I'_w$ if $e_\theta(x)$ has $w$ as prefix. Clearly those intervals
$I'_w$ coincide with the intervals $I_w$ and therefore, the classical
$\theta$-developments are the same as the representation obtained in
the framework of the abstract numeration systems (naturally, under the
extra assumptions of this section corresponding to regular languages
associated to Pisot number).

Now we can use a result of Klaus Schmidt concerning ultimately
periodic $\theta$-developments \cite{KS} and state the following
result.

\begin{The}
  If $L$ is the language of all the representations of the integers in
  a linear Bertrand numeration system associated to a Pisot number
  $\theta$ then the set of real numbers having an ultimately periodic
  representation in the abstract system built upon $L$ is exactly
  $$\mathbb{Q}(\theta)\cap [1/\theta,1].$$
\end{The}

\begin{Rem} 
  In this section, we have only considered the case $e_\theta(1)$
  ultimately periodic. If $e_\theta(1)=t_1\cdots t_m$ is finite
  ($t_m\neq 0$) then the same situation holds. The construction of the
  automaton $\mathcal{A}$ is the same as before but with $N=m$ and
  $p=0$. All the edges from $q_m$ lead to $q_1$ and are labeled by
  $0,\ldots ,t_m-1$. The automaton $\mathcal{A}$ is depicted in Figure
  \ref{fig:efinite}.
  \begin{figure}[htbp]
    \centering
    \includegraphics{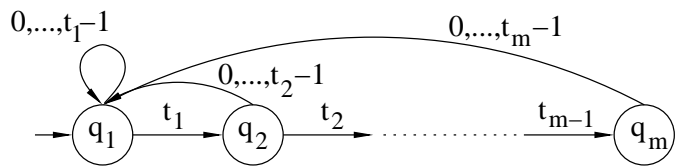}
    \caption{The automaton $\mathcal{A}$ in the case $e_\theta(1)$ finite.}
    \label{fig:efinite}
  \end{figure}
  
\end{Rem}

\end{document}